\documentclass[10pt,twocolumn]{article}

\pdfoutput=1

\usepackage{ifxetex}
\ifxetex
  \usepackage{fontspec}
  \defaultfontfeatures{Ligatures=TeX}
\else
  \usepackage[utf8]{inputenc}
  \usepackage{times}
  \usepackage{inconsolata}
  \usepackage{graphicx}
\fi

\usepackage{url}
\usepackage{listings}
\usepackage[usenames,dvipsnames]{color}

\definecolor{mymauve}{rgb}{0.58,0,0.82}

\lstdefinelanguage{Golo}{
  morekeywords={module,function,let,var,augment,struct,%
  if,else,match,return,%
  case,when,then,otherwise,%
  while,for,foreach,%
  import,try,catch,finally,throw,%
  local,continue,break,%
  is,isnt,oftype,orIfNull,in,or,and,not,null%
  },%
  sensitive,%
  morecomment=[l]\#,%
  morecomment=[s]{----}{----},%
  morestring=[b]",%
  morestring=[b]',%
  showstringspaces=false,%
  commentstyle=\itshape\color{OliveGreen},%
  keywordstyle=\bfseries\color{RedViolet},%
  stringstyle=\color{mymauve},%
  tabsize=2%
}[keywords,comments,strings]

\newcommand{\indy}{\texttt{invokedynamic}~}

\title{Opportunities for a Truffle-based Golo Interpreter}
\author{
  Julien Ponge$^1$,
  Frédéric Le Mouël$^1$,
  Nicolas Stouls$^1$, and
  Yannick Loiseau$^{2}$ \\ \\
  $^1$ Université de Lyon \\
  INSA-Lyon, CITI-INRIA F-69621, Villeurbanne, France \\
  \url{firstname.lastname@insa-lyon.fr} \\ \\
  $^2$ Clermont Université, Université Blaise Pascal, LIMOS, BP 10448, \\
  F-63000, Clermont-Ferrand, France \\
  \url{firstname.lastname@univ-bpclermont.fr}
}
\date{April 2015}

\begin{document}

\maketitle

\begin{abstract}
Golo is a simple dynamically-typed language for the Java Virtual Machine. Initially implemented as a
ahead-of-time compiler to JVM bytecode, it leverages \indy and JSR 292 method handles to implement a
reasonably efficient runtime. Truffle is emerging as a framework for building interpreters for JVM
languages with self-specializing AST nodes. Combined with the Graal compiler, Truffle offers a
simple path towards writing efficient interpreters while keeping the engineering efforts balanced.
The Golo project is interested in experimenting with a Truffle interpreter in the future, as it
would provides interesting comparison elements between invokedynamic \emph{versus} Truffle for
building a language runtime.
\end{abstract}

\section{Introduction}

Golo is a simple dynamically-typed language for the Java Virtual Machine \cite{Ponge2013}. Initially
designed as an experiment around the capabilities of the new \indy JVM instruction that appeared in
Java SE 7 \cite{Rose2009}, it has since emerged as a language supported by a small community that
goes beyond the bounds of academia. Applications have been found in \emph{Internet of Things} (IoT)
settings, and we consider Golo to be small enough to be used for language and runtime experiments by
researchers, students and hobbyists. This claim is supported by examples such as ConGolo, a
derivative experiment for contextual programming\footnote{See
\url{https://github.com/dynamid/contextual-golo-lang}.}, and the community projects\footnote{The
\emph{kiss} web framework is a good example: \url{https://github.com/k33g/kiss}.}. Golo is currently
being proposed for incubation at the Eclipse Foundation in the hope of finding new opportunities and
continuing the development at a vendor-neutral foundation\footnote{See
\url{https://projects.eclipse.org/proposals/golo}.}.

\begin{figure}
\ifxetex
\else
  \footnotesize
\fi
\begin{lstlisting}
module samples.Concurrency

import java.util.concurrent
import gololang.Async

local function fib = |n| {
  if n <= 1 {
    return n
  } else {
    return fib(n - 1) + fib(n - 2)
  }
}

function main = |args| {
  let executor = Executors.newFixedThreadPool(2)
  let results = [30, 34, 35, 38, 39, 40, 41, 42]:
    map(|n| -> executor: enqueue(-> fib(n)):
    map(|res| -> [n, res]))
  reduce(results, "", |acc, next| ->
      acc + next: get(0) + " -> " + next: get(1) + "\n"
  ):
    onSet(|s| -> println("Results:\n" + s)):
    onFail(|e| -> e: printStackTrace())
  executor: shutdown()
  executor: awaitTermination(120_L, TimeUnit.SECONDS())
}

# === Prints the following ===
# Results:
# 30 -> 832040
# 34 -> 5702887
# 35 -> 9227465
# 38 -> 39088169
# 39 -> 63245986
# 40 -> 102334155
# 41 -> 165580141
# 42 -> 267914296
\end{lstlisting}
\caption{Computing Fibonacci numbers in Golo with concurrent and asynchronous APIs.}
\label{fig:sample}
\end{figure}

Figure~\ref{fig:sample} provides a sample Golo program. It computes several Fibonacci numbers with
the naive recursive definition of the \texttt{fib} function. It takes advantage of regular Java
executors and Golo APIs for promises and futures \cite{Liskov1988} to perform the computations on 2
worker threads, and collect the results through reduction of futures.

Briefly, the main characteristics of the Golo programming language are the following:
\begin{itemize}
  \item dynamic typing using Java types,
  \item higher-order functions and binding to Java single-method and functional interfaces,
  \item ability to \emph{augment} existing types (including from JVM languages) with new methods,
  \item tuples and structures (augmenting the later is reminiscent of Go-style objects \cite{Pike2012}),
  \item dynamic objects with instance-level definitions,
  \item Python-style decorators (i.e., higher-order function-based).
\end{itemize}

Unlike many other JVM languages such as JRuby, Jython or Nashorn, Golo is not a port of an
existing language to the JVM and invokedynamic. This is interesting, as Golo was designed
\emph{around} the capabilities of invokedynamic, which gives a different perspective on the design
of a invokedynamic-based runtime.

\section{Ahead-of-time compilation based on JSR 292}

Golo uses ahead-of-time bytecode generation rather than interpretation. The grammar of Golo is
written using the $LL(k)$ JJTree / JavaCC parser generator \cite{Kodaganallur04}, mainly due to its
simplicity and lack of a runtime dependency, as it generates all the Java code required for a
working parser. The front-end generates an abstract syntax directly from JJTree, which is then
transformed into an intermediate representation based on a Golo-specific object model, comprising
classes to model reference lookups, functions, common statements and so on. The intermediate
representation is visited by several phases to check for undeclared references, expand lambda
functions / closures to anonymous functions, and ultimately generated JVM bytecode with the popular
ASM library \cite{Bruneton02}.

\paragraph{Stable bytecode, adaptive runtime dispatch.}
The compiler generates a largely \emph{untyped} bytecode. Most references are on the
\texttt{java.lang.Object} type, with some peculiar portions of the bytecode doing cast checks (e.g.,
branch conditions require refining to \texttt{java.lang.Boolean} and unboxing the primitive
\texttt{boolean} value). The generated bytecode remains stable at runtime, unlike speculations and
invalidations as found in Nashorn to try to take advantage of primitive types when
possible.

\begin{figure*}[t]
  \centering
  \includegraphics[width=0.8\textwidth]{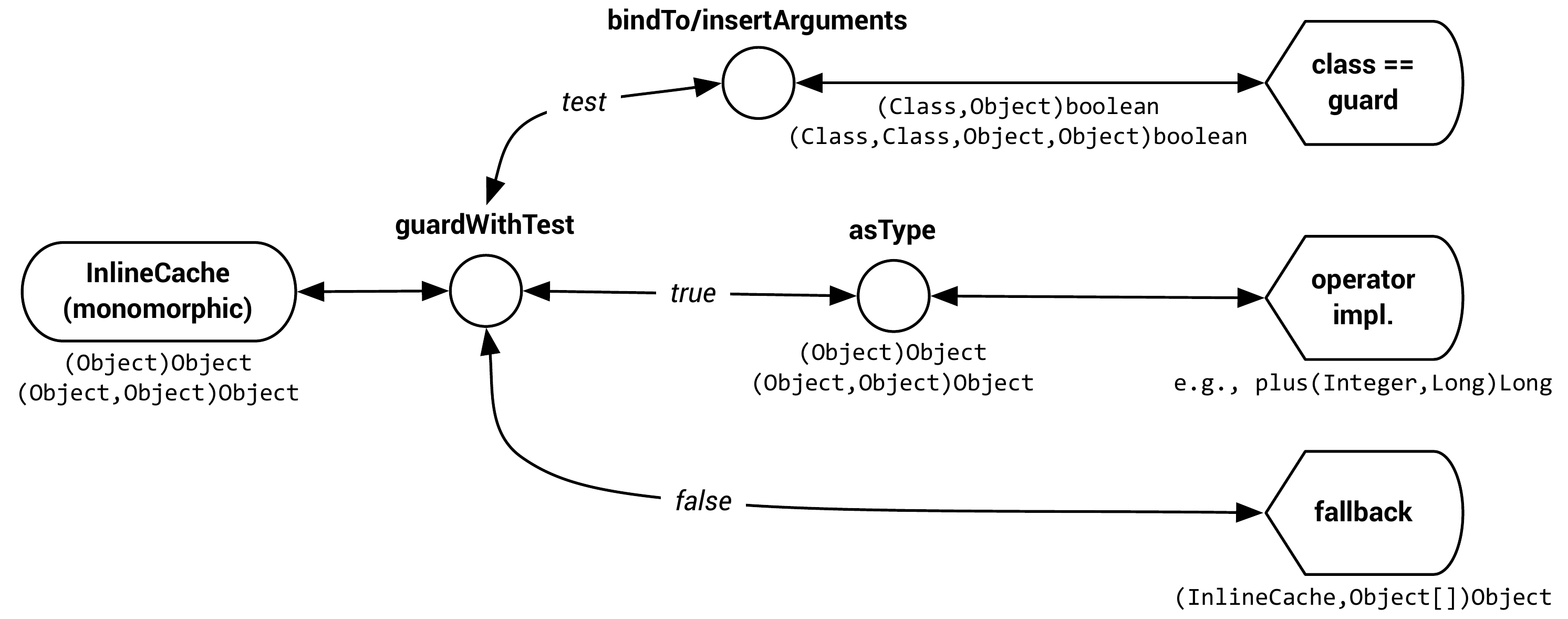}
  \caption{Operator monomorphic inline-cache based on method handles.}
  \label{fig:callsite}
\end{figure*}

As most call sites (including arithmetic operators) are based on invokedynamic, the runtime adapts
the dispatch targets through evolving \emph{method handle} chains, based on types observed at
runtime. Figure~\ref{fig:callsite} gives an example: operators use a monomorphic inline-cache
construct \cite{Holzle91}. The construction relies on a \emph{guarded} combinator that dispatches to
the right target as long as type remain stable (e.g., \texttt{plus(Integer, Long) Long} for
\texttt{10 + 10\_L}). The fallback branch points to a handler that dynamically finds a new target
based on the observed types, and overwrites the call site method handle dispatch chain with the new
one.

\begin{figure}[htb]
  \includegraphics[width=\columnwidth]{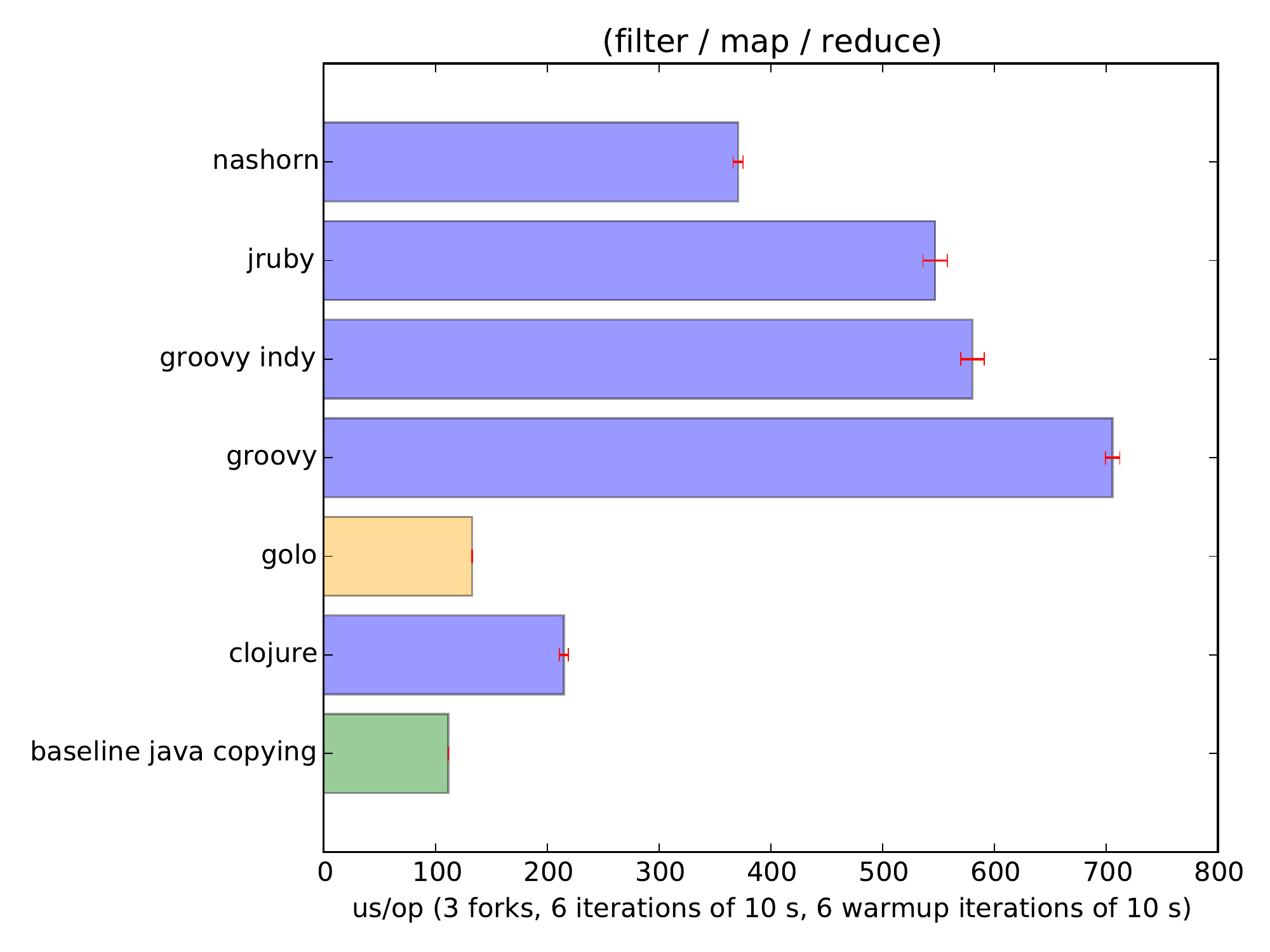}
  \caption{Filter / Map / Reduce micro-benchmark.}
  \label{fig:filter-map-reduce}
\end{figure}

\begin{figure}[htb]
  \includegraphics[width=\columnwidth]{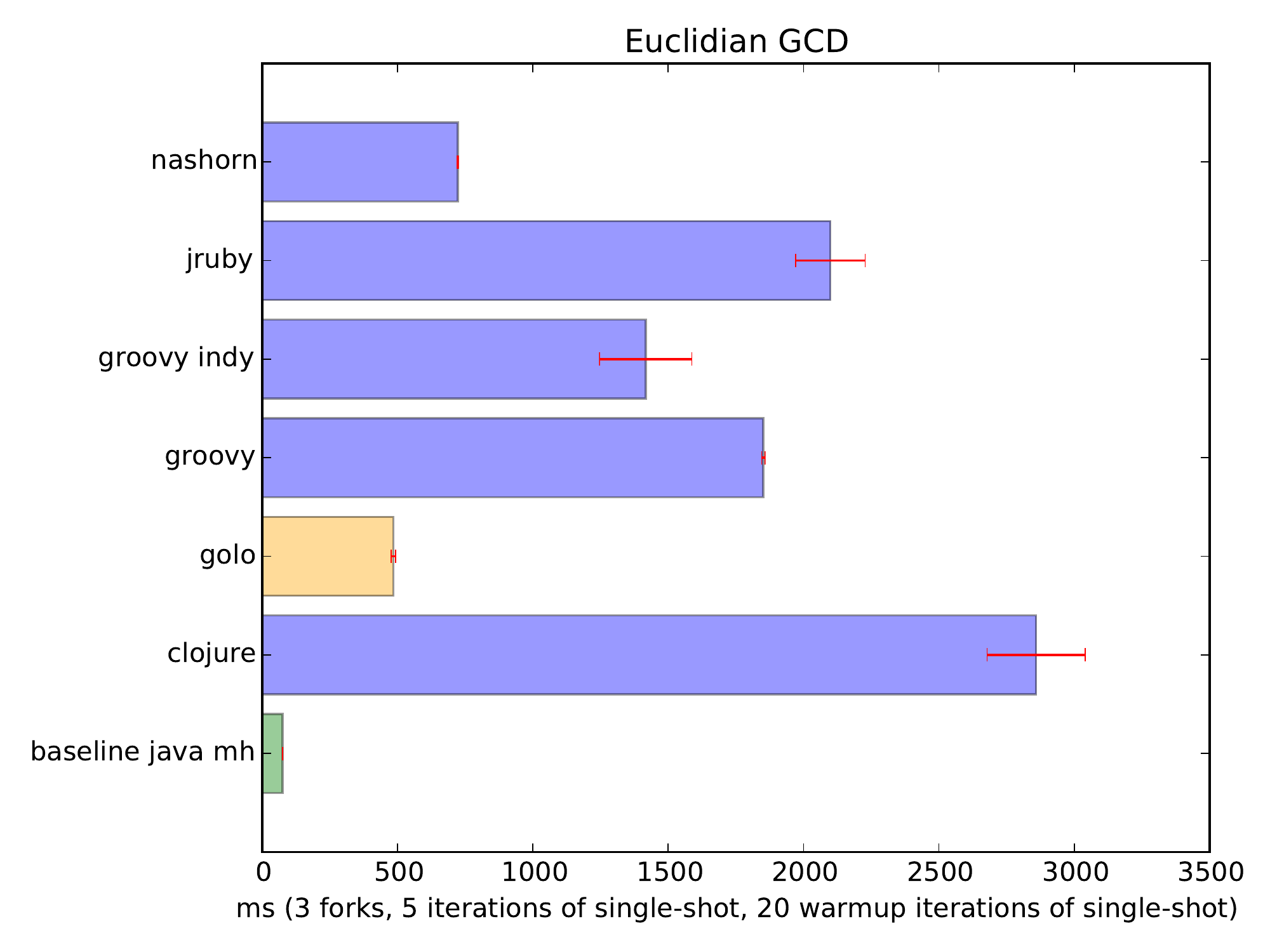}
  \caption{Greatest common divisor micro-benchmark.}
  \label{fig:gcd}
\end{figure}

\paragraph{Performance considerations.}
In general, Golo exhibits good performance on function and method dispatch\footnote{See \url{https://github.com/golo-lang/golo-jmh-benchmarks} for a collection of micro-benchmarks.}.
Figure~\ref{fig:filter-map-reduce}  shows the results of a micro-benchmark based on applying the
usual \emph{filter}, \emph{map} and \emph{reduce} operations on collections. Golo is practically as
fast as a baseline in Java where the operations are implemented using collection
copies\footnote{This micro-benchmark was written while Golo was still compatible with Java SE 7,
hence it would be interesting to compare with Java SE 8 streams.}.

Figure~\ref{fig:gcd} shows a GCD micro-benchmark. While performance remains good compared to other
dynamic languages, it highlights the performance bottleneck due to boxing of primitive types, which
is also further confirmed by further nano-benchmarks that we have. We are planning to explore ways
to be clever than we are at the moment with respect to arithmetic operations.

\section{Perspectives with Truffle}

Writing an interpreter for Golo based on Truffle \cite{marr14} is interesting for comparing the
effectiveness of invokedynamic \emph{versus} Truffle to implement common language runtime patterns
(e.g., arithmetic operations or inline-caches). In terms of performance, the following points are of
comparison interest:
\begin{enumerate}
  \item functions and methods dispatch,
  \item arithmetic operations (Truffle node specialisation can potentially eliminate some boxings),
  \item dispatch in Golo dynamic objects (Truffle proposes a \emph{Shapes} abstraction),
  \item statistical optimizations for application profiling (Truffle exposes node counters that
        could be use to mine application behavior and dynamically activate relevant optimizations).
\end{enumerate}

We are looking forward to experimentions with Truffle in the near future, and have elements of
comparisons in the challenges of designing programming language on the JVM.

\bibliographystyle{unsrt}
\bibliography{biblio}

\end{document}